\documentclass[10pt,conference]{IEEEtran}

\usepackage{cite}
\usepackage{amsmath,amssymb,amsfonts}
\usepackage{algorithmic}
\usepackage{graphicx}
\usepackage{textcomp}
\usepackage{xcolor}
\usepackage{booktabs}
\usepackage{url}
\usepackage{listings}
\usepackage{tikz}
\usepackage{pgfplots}
\pgfplotsset{compat=1.17}
\usepgfplotslibrary{groupplots}
\usepackage{flushend} 
\usetikzlibrary{positioning,arrows.meta,shapes,fit,backgrounds}

\lstdefinelanguage{Coq}{
  morekeywords={Definition, Proof, Qed, Lemma, Theorem, intros, apply,
    destruct, split, done, exists, move, first, by, try},
  sensitive=true,
  morecomment=[l]{(*},
  morecomment=[s]{(*}{*)},
  morestring=[b]",
}

\lstdefinelanguage{HHL}{
  morekeywords={workflow, workspace, hook, invoke, let, var, if, else,
    for, while, in, match, print, call, python, import, return, true, false,
    string, bool, int, unit, list, set, option,
    Write, Edit, Bash, Read, Turn,
    Allow, Deny, Ask, Ok, Err, Some, None,
    pre, post, cwd, system_prompt, allowed_tools, model, max_turns},
  sensitive=true,
  morecomment=[l]{//},
  morecomment=[s]{/*}{*/},
  morestring=[b]",
  alsoletter={\_},
}

\lstdefinestyle{chlstyle}{
  language=HHL,
  basicstyle=\footnotesize\ttfamily,
  keywordstyle=\bfseries\color{blue!70!black},
  commentstyle=\itshape\color{gray},
  stringstyle=\color{green!50!black},
  frame=single,
  xleftmargin=2em,
  framexleftmargin=1.5em,
  breaklines=true,
  columns=flexible,
  showstringspaces=false,
}

\lstdefinestyle{coqstyle}{
  language=Coq,
  basicstyle=\small\ttfamily,
  keywordstyle=\bfseries,
  commentstyle=\itshape\color{gray},
  frame=single,
  xleftmargin=2em,
  framexleftmargin=1.5em,
  breaklines=true,
  columns=flexible,
  escapechar=@,
}

\begin{document}

\bstctlcite{IEEEexample:BSTcontrol}

\title{Harnessing Code Agents for Automatic Software Verification}

\author{
\IEEEauthorblockN{Shuangxiang Kan}
\IEEEauthorblockA{\textit{Singapore Management University} \\
Singapore \\
sxkan@smu.edu.sg}
\and
\IEEEauthorblockN{Shuanglong Kan}
\IEEEauthorblockA{\textit{Barkhausen Institut} \\
Dresden, Germany \\
shuanglong.kan@barkhauseninstitut.org}
\and
\IEEEauthorblockN{Sebastian Ertel}
\IEEEauthorblockA{\textit{Barkhausen Institut} \\
Dresden, Germany \\
Sebastian.Ertel@barkhauseninstitut.org}
}

\maketitle
\thispagestyle{plain}
\pagestyle{plain}

\begin{abstract}
Formal verification offers the strongest available guarantee of
software correctness, but it does not scale: the proofs demanded by
interactive theorem provers such as Coq require enormous expert effort.
Large language
models (LLMs) promise to generate these proofs automatically. Existing
approaches wire a fixed, human-designed proof strategy into the system
and constrain the model to follow it---retrieving relevant premises and
predicting tactics one step at a time, or recursively splitting a goal
into subgoals by divide-and-conquer---yet still prove only a
fraction of their target theorems.

We show that imposing such a strategy is unnecessary, and limiting:
handing the whole lemma to a general LLM \emph{code agent}---for
example, Claude Code~\cite{ClaudeCodeSDK}---free to choose its own
approach, and wrapping
it in a \emph{verification harness} is both simpler and more effective,
achieving surprisingly \emph{full} coverage---every targeted lemma proved,
with no failures and no Coq expert intervention. The agent writes the proofs
with effective feedback and hard constraints from the harness that keep
each one \emph{sound} (accepted only when the prover's kernel closes it),
\emph{complete} (no obligation left unproved or silently dropped), and
\emph{terminating} (no divergent, non-terminating tactics).
We evaluate the harness + code agent for verified software development
along three dimensions.
(1)~\emph{Core logic for software verification:} on Iris, the state-of-the-art
separation logic for concurrent and memory-manipulating programs, Aria
proves all 4{,}257 lemmas of the four
core modules and the 217 lemmas verifying Rust's standard libraries
(\texttt{Arc}, \texttt{Mutex}, \texttt{RwLock}, \texttt{RefCell}) built on
it---both in full and fully automatically. (2)~\emph{Comparison with
prior LLM provers:} on reglang, where prior LLM provers manage barely one in
eight, Aria proves all 318. (3)~\emph{Generality across provers:} on iris-lean,
the unfinished Lean~4 port of Iris, it proves 72 not-yet-ported lemmas,
showing the approach is not specific to Coq. We
conclude that a state-of-the-art model---Claude Opus~4.7---is capable of
writing proofs for state-of-the-art verified software development
\emph{fully and automatically}.
\end{abstract}


\section{Introduction}
\label{sec:introduction}

Software defects cost the global economy trillions of dollars each year,
and the systems where a single bug is most catastrophic---compilers,
operating-system kernels, cryptographic libraries---are precisely the
ones that most need strong correctness guarantees. Formal verification
provides the strongest such guarantee available: a machine-checked
proof, constructed in an interactive theorem prover (ITP) such as
Coq~\cite{Coq2004, BertotCasteran2004} or
Isabelle~\cite{Nipkow2002Isabelle}, that a program meets its
specification. Landmark efforts such as the CompCert verified compiler (developed roughly 20 years as of 2026)
demonstrate that this level of assurance is attainable in practice. Yet
verified software remains the exception rather than the rule, for one
stubborn reason: the proofs must be written by hand.

The obstacle is the cost of those proofs. Constructing an ITP proof
requires a highly trained expert to compose tactics one step at a time, reasoning
simultaneously about the program, its specification, and the prover's
underlying type theory. Nowhere is this harder than in separation
logic~\cite{OHearn2001} and its modern realization,
Iris~\cite{Jung2015, Jung2018}, where concurrency, higher-order state,
and \emph{ghost resources}---auxiliary, proof-only bookkeeping, invisible
to the running program, that tracks how shared state is allowed to
evolve---can make a single lemma take an expert hours or even days to
discharge. This manual-proof bottleneck---more than any
limitation of the underlying theory---is what keeps verified software
from scaling to everyday engineering practice.

A growing body of work attacks this bottleneck with machine learning.
Early systems learned to predict Coq tactics from corpora of human
proofs and drove a search over the proof
tree~\cite{Yang2019CoqGym, SanchezStern2020Proverbot,
Blaauwbroek2020Tactician, Blaauwbroek2024Graph2Tac}. Large language
models (LLMs) then shifted the paradigm toward generating whole proofs
or large fragments, augmented with premise retrieval and feedback-driven
repair, with strong results across
Isabelle~\cite{First2023Baldur, Jiang2023DSP},
Lean~\cite{Yang2023LeanDojo, Xin2024DeepSeekProver},
Metamath~\cite{Polu2020GPTf}, and
Coq~\cite{Lu2024PALM, Thompson2025Rango, Thakur2024COPRA}.

Two limitations recur across this prior work, summarized for the Coq
systems in Table~\ref{tab:priorcoq}. First, although
these systems are automatic, their \textit{coverage} is partial:
reported success rates span roughly 12\%--48\% of the theorems
attempted---PALM proves 40.4\% of 10{,}842 CoqGym
theorems~\cite{Lu2024PALM}, Rango 32.0\% of 10{,}396 CoqStoq
theorems~\cite{Thompson2025Rango}, and the strongest LLM agent, COPRA,
48.3\% of a small 118-theorem CompCert subset~\cite{Thakur2024COPRA}.
Second---and more telling---this partial coverage is measured on the
\textit{easy} part of the problem. The benchmarks are drawn from real Coq
developments but consist almost entirely of \textit{sequential} program
and datatype verification; they omit the two dimensions that make
real-world software verification hard: \textit{concurrency} and
fine-grained reasoning about \textit{shared, mutable memory}---exactly
what separation logic exists to handle. CompCert is a sequential C
compiler, and CoqGym and CoqStoq aggregate open-source projects with
little concurrent or heap-intensive code. The higher-order, concurrent
separation logic that Iris embodies---where interacting threads and the
shared state they manipulate make proofs dramatically harder---is left
untouched.

\begin{table}[t]
\centering
\caption{Learning- and LLM-based automated proof systems for Coq.}
\label{tab:priorcoq}
\footnotesize
\begin{tabular}{l r l r r}
\toprule
\textbf{System} & \textbf{Year} & \textbf{Benchmark} & \textbf{Theorems} & \textbf{Proved} \\
\midrule
ASTactic~\cite{Yang2019CoqGym}        & 2019 & CoqGym           & 13{,}137 & 12.2\%$^{\ddagger}$ \\
Proverbot9001~\cite{SanchezStern2020Proverbot} & 2020 & CompCert & 501 & 19.4\%$^{\ddagger}$ \\
Graph2Tac~\cite{Blaauwbroek2024Graph2Tac} & 2024 & Coq Opam pkgs  & 2{,}000$^{\S}$  & 26.1\%$^{\S}$ \\
PALM~\cite{Lu2024PALM}                & 2024 & CoqGym           & 10{,}842 & 40.4\% \\
COPRA~\cite{Thakur2024COPRA}          & 2024 & CompCert$^{\dagger}$ & 118  & 48.3\% \\
Rango~\cite{Thompson2025Rango}        & 2025 & CoqStoq          & 10{,}396 & 32.0\% \\
Cobblestone~\cite{Kasibatla2026Cobblestone} & 2026 & CoqGym & 100$^{\P}$ & 48\% \\
ProofCoop~\cite{Kaufman2026ProofCoop} & 2026 & CoqGym & $\sim$11.9K & 33.0\%$^{\ddagger}$ \\
\bottomrule
\end{tabular}

\smallskip
{\footnotesize
$^{\dagger}$118-theorem subset of CompCert (501).
$^{\ddagger}$Learned model alone, no hammer/solver.
$^{\S}$Random 2{,}000-theorem sample; $33.2\%$ with a $k$-NN solver.
$^{\P}$One of four 100-theorem subsets; $55$--$58\%$ with oracle guidance.}
\end{table}

Despite their variety, these systems share a commitment: each wraps the
model in a hand-engineered strategy that dictates \textit{how} a proof is
built. One family works tactic by tactic, searching the proof
tree---a trained tactic predictor~\cite{Yang2019CoqGym,
SanchezStern2020Proverbot, Blaauwbroek2020Tactician,
Blaauwbroek2024Graph2Tac}, an in-context LLM with
backtracking~\cite{Thakur2024COPRA}, or retrieval-augmented step
prediction~\cite{Thompson2025Rango}. Another generates a whole proof and
wraps it in an engineered loop---symbolic repair with
backtracking~\cite{Lu2024PALM}, or divide-and-conquer over
subgoals~\cite{Kasibatla2026Cobblestone}. Either way, a fixed,
human-designed procedure governs the construction.

This paper shows that the scaffolding is unnecessary: general \textit{code
agents} such as Claude Code and Codex can replace it with better results.
In confining the model to a narrow role---predict a tactic, or fill a slot
in a repair loop---prior systems never tap the power of such an agent: one
that reads the context, decides what to do next, and builds the proof
however it sees fit. Our method is
deliberately coarse: we hand a whole lemma to a general LLM \textit{code
agent}---an off-the-shelf coding tool built for everyday software
engineering, here Claude Code on Claude Opus~4.7, not specialized for Coq
or theorem proving---and let it write a complete proof however it sees fit.
We drive the agent non-interactively through Claude Code's \texttt{claude -p}
command, which embeds in an ordinary program (a Python driver, in our
case), so the whole campaign runs automatically with no human at the
keyboard.

The agent is free to err: a \textit{verification harness} built on the Coq
kernel checks each attempt and, on rejection, returns precise feedback (the
failing step and the open goal); the agent revises and retries the whole
proof, up to thirty times, until the kernel accepts it. We impose no premise
selector, tactic-level search, decomposition, or repair procedure---only
the verifier and a thin policy around it. Surprisingly, this is enough,
and not merely on easy benchmarks: it reaches \textit{full} coverage on
the hardest separation-logic proofs in existence. Two properties of
interactive theorem proving make it work. \textit{Trust is free}: the
kernel rejects anything incorrect, so the confident-but-wrong output that
makes LLMs risky elsewhere~\cite{LLMHallucination2023} cannot slip
through---the difficulty is never whether to believe a proof, only whether
the model can \textit{find} one, a long, exact, multi-step construction in
which a single wrong step is fatal. And \textit{the harness closes the
loop}: rather than produce a correct proof in one shot, the agent proposes
a candidate, reads the exact error, and revises, turning proof
construction into a guided search against ground truth that a capable
agent can direct on its own.

We realize this idea as \textit{Aria}. To make the harness reusable, we
express it in a declarative language, the \textit{Harness Hook
Language} (HHL), written against a general code-agent interface so that
the same harness can drive different code agents, such as Claude Code and
Codex; in this paper it targets the Claude Code SDK~\cite{ClaudeCodeSDK}. A harness written in HHL bundles
the checks that surround the agent for proof writing: a \textit{timeout} check for divergent,
non-terminating tactics; a \textit{hallucination} check that rejects
proofs closed with \texttt{Admitted} or with target lemmas silently
dropped; an \textit{Iris linter}; and final
\textit{Coq kernel} verification. Each lemma runs in a single autonomous
session---the agent attempts a proof, the harness checks it, and on
failure the agent retries on the returned feedback---with no human in the
loop.

We evaluate Aria on four bodies of proofs---three in Coq and one in
Lean. Its main target is the
Iris development~\cite{IrisDocumentation, Krebbers2014}---the
state-of-the-art separation logic for verifying concurrent,
memory-manipulating programs, and arguably the most demanding such proofs
in existence, a real, curated artifact rather than a synthetic benchmark. Aria proves \textbf{all 4{,}257
lemmas} in Iris's four core modules
(\texttt{algebra}, \texttt{bi}, \texttt{base\_logic},
\texttt{program\_logic}) fully automatically, with zero expert
intervention and zero failures. Three further settings then test whether
this result holds beyond the Iris core. The first is the verified
software built \emph{on} Iris: the RustBelt safety
proofs~\cite{Jung2018RustBelt} for Rust's standard libraries---its
concurrency and interior-mutability types such as
\texttt{Arc}, \texttt{Rc} and
\texttt{Weak}, \texttt{Mutex} and \texttt{RwLock}, \texttt{RefCell}, and
\texttt{Cell}; Aria successfully proves \textbf{all 217 lemmas}. The second
is a direct comparison with prior work on its
own terms: reglang, a 318-theorem library of regular-language
theory. Far less demanding
than the Iris core, reglang is nonetheless the CoqStoq benchmark on
which prior LLM provers fare \emph{worst}, proving barely one in
eight~\cite{Thompson2025Rango}; Aria successfully proves them all. The
third moves to a different prover entirely: iris-lean, the in-progress
Lean~4 port of Iris, where we target lemmas not yet ported, and Aria proves
all \textbf{72}.
Aria is complete across all four settings.

This paper makes the following contributions:
\begin{itemize}
  \item \textit{Harness + code agent.} We propose pairing a
    general-purpose LLM code agent with a verification harness, imposing
    no Coq-specific proof strategy, and show that this simple design
    surprisingly achieves \textit{full} coverage---every targeted lemma
    proved, with no failures and no Coq expert intervention.
  \item \textit{Evaluation on Iris and its Rust applications.} We
    evaluate the approach on Iris, the state-of-the-art higher-order
    concurrent separation logic, together with the Rust library verification
    built on it, proving both completely. This is, to our knowledge, the
    first empirical evidence that a current state-of-the-art LLM can write
    the proofs of an expert-grade, state-of-the-art software-verification
    Coq library.
  \item \textit{HHL: a language for writing harnesses around LLM agents.}
    We introduce HHL, a declarative language that makes the harness a
    reusable, auditable artifact---written once and applied across lemmas,
    libraries, and agent backends---rather than ad-hoc glue code.
\end{itemize}

The rest of the paper covers the Iris framework
(Section~\ref{sec:iris}), Aria's architecture
(Section~\ref{sec:architecture}) and the HHL language
(Section~\ref{sec:language-design}), its implementation
(Section~\ref{sec:implementation}), the evaluation
(Section~\ref{sec:experiment}), and discussion and related work
(Sections~\ref{sec:discussion} and~\ref{sec:related}).

\section{The Iris Separation Logic Framework}
\label{sec:iris}

In this section we introduce Iris, the primary benchmark of this work. The
premise that motivates the whole paper is this: \textit{if an automated
system can master proofs as hard as Iris's, then---because so much software
verification is built on Iris---it removes the manual-proof bottleneck not for Iris
alone but across that whole body of software, letting \textbf{verified
software development} scale to everyday engineering practice.} The rest of
this section makes the case for that premise: we sketch what Iris is and
how it grew, survey the verified software built on it, and explain why its
proofs are so hard---the difficulty that motivates automating them.

Iris~\cite{Jung2015,Jung2018} is a framework for proving that
concurrent, memory-manipulating programs meet their specifications. It is
the modern realization of separation logic~\cite{OHearn2001}, whose
central idea---reasoning about disjoint regions of memory
independently---makes verification of pointer-heavy code tractable.
Developed at MPI-SWS and Aarhus University and first released in 2015, Iris
has been refined for more than a decade into a large, actively maintained
Coq library, extending separation logic to handle the features pervasive
in real-world software: concurrency, higher-order functions, and the
fine-grained sharing of mutable memory.

Iris is not a single-purpose tool but the substrate beneath a broad and
growing body of verified software. Its best-known application is
RustBelt~\cite{Jung2018RustBelt}, the first formal safety proof for the
core of Rust, showing that Rust's type system safely encapsulates the
\texttt{unsafe} code in its standard library. Beyond it, Iris underpins
verified work across many domains: programming-language safety
(RustHornBelt~\cite{Matsushita2022RustHornBelt},
RefinedRust~\cite{Gaher2024RefinedRust}, Rust under relaxed
memory~\cite{Dang2020RustBeltRelaxed}, Iris-Wasm~\cite{Rao2023IrisWasm});
systems software such as C verification
(RefinedC~\cite{Sammler2021RefinedC}, Quiver~\cite{Spies2024Quiver}),
crash-safe storage in Go (Perennial~\cite{Chajed2019Perennial},
GoJournal~\cite{Chajed2021GoJournal}, DaisyNFS~\cite{Chajed2022DaisyNFS}),
and hypervisors~\cite{Liu2023VMSL}; distributed and concurrent systems
(Aneris~\cite{KroghJespersen2020Aneris}, Grove~\cite{Sharma2023Grove},
Trillium~\cite{Timany2024Trillium}, Actris~\cite{Hinrichsen2020Actris},
Hazel~\cite{DeVilhena2021Hazel}, Cosmo~\cite{Mevel2020Cosmo},
Zoo~\cite{Allain2026Zoo}); compilers and machine code
(Simuliris~\cite{Gaher2022Simuliris}, Islaris~\cite{Sammler2022Islaris},
Cerise~\cite{Georges2024Cerise}, Melocoton~\cite{Gueneau2023Melocoton},
DimSum~\cite{Sammler2023DimSum}); and security
(SeLoC~\cite{Frumin2021SeLoC}, Cryptis~\cite{AzevedoDeAmorim2026Cryptis}).
This breadth places Iris at the heart of modern \textbf{verified
software development}: it sits under programming languages,
compilers, distributed systems, and machine code alike.

That reach comes at a cost. Iris proofs are among the hardest in all of
software verification: a single non-trivial lemma can take an expert hours
or even days, because the proof must account for every way concurrent
threads can interleave and share state, with little syntactic guidance on
what step to take next~\cite{Krebbers2014, IrisDocumentation}. Because Iris
demands such deep understanding, developing verified software on top of it
requires experts who have that understanding---a bottleneck that keeps
verified software from scaling. That is what makes the premise above worth
testing: automate proofs this hard, and the bottleneck eases for the broad
range of software built on Iris.

\section{Architecture}
\label{sec:architecture}

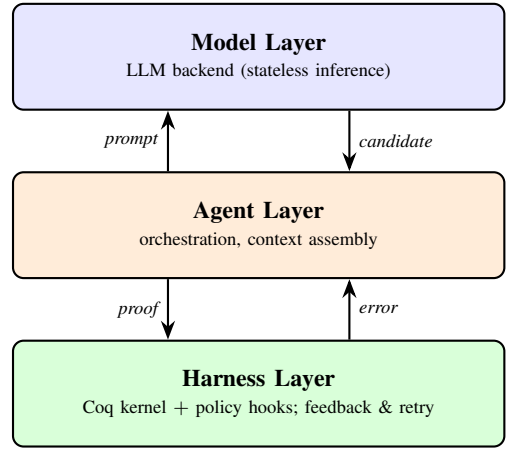
\begin{figure}[t]
\centering
\begin{tikzpicture}[
  node distance=8mm,
  layer/.style={
    rectangle, rounded corners, draw, thick,
    minimum width=6.5cm, minimum height=1.4cm,
    align=center, font=\small
  },
  modelL/.style={layer, fill=blue!10},
  agentL/.style={layer, fill=orange!15},
  harnessL/.style={layer, fill=green!15},
  arr/.style={-{Stealth[length=2.5mm]}, thick},
  lbl/.style={font=\scriptsize\itshape, midway}
]

\node[modelL] (model) {%
  \textbf{Model Layer}\\
  {\scriptsize LLM backend (stateless inference)}};
\node[agentL, below=of model] (agent) {%
  \textbf{Agent Layer}\\
  {\scriptsize orchestration, context assembly}};
\node[harnessL, below=of agent] (harness) {%
  \textbf{Harness Layer}\\
  {\scriptsize Coq kernel $+$ policy hooks; feedback \& retry}};

\draw[arr] ([xshift=-1.2cm]agent.north) -- ([xshift=-1.2cm]model.south)
  node[lbl, left] {prompt};
\draw[arr] ([xshift=1.2cm]model.south) -- ([xshift=1.2cm]agent.north)
  node[lbl, right] {candidate};

\draw[arr] ([xshift=-1.2cm]agent.south) -- ([xshift=-1.2cm]harness.north)
  node[lbl, left] {proof};
\draw[arr] ([xshift=1.2cm]harness.north) -- ([xshift=1.2cm]agent.south)
  node[lbl, right] {error};

\end{tikzpicture}
\caption{Three-layer architecture of Aria.}
\label{fig:architecture}
\end{figure}

The Aria system, our system for proving Coq lemmas fully automatically
with a code agent, is organized as a three-layer architecture:
the \textit{model layer}, the \textit{agent layer}, and the
\textit{harness layer}. This separation cleanly decouples the
non-deterministic reasoning capabilities of the LLM from the
deterministic correctness guarantees of the proof assistant.
Figure~\ref{fig:architecture}
illustrates the overall structure and the information flow between
layers.

\subsection{Model Layer}
\label{sec:arch:model}

The model layer is the source of proof candidates. It consists of one or
more LLMs---for example, proprietary models from Anthropic, Google
Gemini, and OpenAI, or open-source models such as Llama, Qwen, and
DeepSeek---accessed through a uniform inference interface that treats each
model as an interchangeable backend. Given a structured prompt that
encodes the current proof obligation together with the surrounding
context and any prior feedback, the model generates a candidate
proof and writes it into the
source file at the appropriate location.

\subsection{Agent Layer}
\label{sec:arch:agent}

The agent layer is the orchestrator and the locus of intelligence
beyond raw model inference. It owns both proof construction and strategy:
choosing which lemma to attack and how to approach it---retrieving similar
proofs, definitions, dependencies, and prior lemmas as context---then
prompting the model and interpreting its response into a candidate proof.
When the harness returns an error and the pending goals of a proof, the agent revises
accordingly; the retry loop itself is driven by the harness layer below. The agent layer also enforces the session-scoping policy,
ensuring that context remains
focused and token usage bounded.

The agent layer is not tied to a particular code agent: any runtime
that exposes the interfaces our harness relies on---tool
dispatch and the pre-/post-tool and turn-level \textit{hooks} through
which policy is enforced---can serve as the agent layer, whether Claude
Code, Codex, or another coding agent.

\subsection{Harness Layer}
\label{sec:arch:harness}

The harness layer is the trust anchor of the system. At its core it wraps
the Coq proof assistant as a verification oracle, but it also runs the
auxiliary policy checks that surround it---a per-tactic timeout for
divergent tactics, a ban on \texttt{admit}/\texttt{Admitted} (which
indicate dropped proofs), a lemma-coverage check, an Iris linter, and
shell-command safety. Every
candidate the agent produces must pass through the harness before it
counts as successful: the harness submits the proof to Coq and, on
rejection, returns a structured report---the failing step, the error
message, and the pending goal---and drives the feedback loop, handing that
report back to the agent and re-dispatching until a proof is accepted or
the retry budget is exhausted. The harness does more than judge
correctness. Correctness is decided by the Coq kernel, not the model or the
agent, so no unverified proof can leave the system. Beyond that, the
harness governs \emph{style}, holding each proof to Iris's conventions, and
\emph{safety}, blocking unwanted or destructive shell commands. This makes
the harness the foundation upon which the entire LLM~+~Harness paradigm
rests.

\section{The Harness Hook Language (HHL)}
\label{sec:language-design}

A central design choice of Aria is to expose the harness layer
through a dedicated, declarative language rather than as an ad-hoc
library of glue code. We call this language HHL. HHL is the surface
through which proof engineers
describe \textit{how} the LLM should be driven, \textit{what} actions
it is allowed to take, and \textit{which} verification steps must
succeed before a proof is accepted. By making these concerns
first-class language constructs, HHL keeps the agent layer concise and
auditable while letting the harness layer enforce strong correctness
guarantees uniformly across all proof attempts.

\subsection{Overview by Example}
\label{sec:chl:overview}

\begin{figure}[t]
\caption{An example HHL program.}
\label{lst:hhl}
\vspace{4pt}
\begin{lstlisting}[style=chlstyle]
hook check_imports: Write|Edit(file, original)
    = check_imports_fn
hook no_banned: Write|Edit(file, original)
    = no_banned_fn
hook safe_command:  Bash(cmd) = safe_command_fn
hook iris_lint:     Turn(files) = iris_lint_fn
hook verify_proof:  Turn(files) = verify_proof_fn
workspace coq_project {
  cwd           = "/path/to/project"
  system_prompt = "You are an Iris expert ..."
  allowed_tools = ["Read", "Bash", "Write", "Edit"]
  max_turns     = 10
}
workflow prove_lemmas(lemmas: list[string]) {
  for lemma in lemmas {
    var proved   = false
    var tries    = 0
    var feedback = ""
    while !proved && tries < 30 {
      let r = invoke(coq_project,
        "Prove the lemma " ++ lemma ++ feedback,
        pre:[check_imports,no_banned,safe_command],
        post:[iris_lint, verify_proof])
      proved   = !r.is_error
      feedback = "\nPrevious error: " ++ r.result
      tries    = tries + 1
    }
    if proved { print("Proved: " ++ lemma) }
    else { print("Failed: " ++ lemma) }
  }
}
\end{lstlisting}
\end{figure}

We introduce HHL by walking through the example program in
Figure~\ref{lst:hhl} (a simplified version of Aria), whose
\texttt{prove\_lemmas} workflow proves a list of Iris lemmas under a policy
of five hooks. An HHL program begins with its
\texttt{hook} declarations: each binds a tool event to a Python policy
function that runs around the corresponding agent action, bracketing it:
\begin{center}
\textit{pre-hooks} $\;\longrightarrow\;$ \textbf{agent action}
$\;\longrightarrow\;$ \textit{post-hooks}
\end{center}
They fire in list order: each pre-hook runs (and may block the action)
before it, left to right, and each post-hook runs after it, in turn.

Each declaration, \texttt{hook}~\textit{name}\texttt{:}~%
\textit{Event}\texttt{(}\textit{payload}\texttt{)} \texttt{=} \textit{fn},
binds a Python function to a tool event. A \textit{pre-hook} such as
\texttt{hook no\_banned: Write|Edit(file, original) = no\_banned\_fn} fires
before a \texttt{Write} or \texttt{Edit} tool call, inspects the payload (here the \texttt{file} being
written and its \texttt{original} contents, from which it can compute the
edit's diff) and returns \texttt{Allow} or \texttt{Deny} to permit or block
it---here denying the edit whenever its diff introduces a banned construct
such as \texttt{admit} or \texttt{Admitted}, or anything else the policy
forbids, such as an \texttt{Axiom} that assumes the goal outright. A
\textit{post-hook} instead runs after a tool call---or, for the
\texttt{Turn} event, at the end of a turn (one round of model output and
the tool calls it triggers)---and returns \texttt{Ok} or
\texttt{Err} to accept or reject the attempt; \texttt{verify\_proof}, for
example, type-checks the edited file with the Coq kernel, which can only be
done once the edits are in place. Besides \texttt{no\_banned} above, the
two remaining pre-hooks are:
\begin{itemize}
  \item \texttt{check\_imports} --- rejects any edit that adds a new
    \texttt{Require} or \texttt{Import}, so a proof may use only the
    modules already in scope; for our experiments, the existing imports are
    sufficient;
  \item \texttt{safe\_command} --- screens shell commands, blocking build
    and file-deleting (\texttt{rm}) commands.
\end{itemize}
Build and verification commands are blocked for two reasons: the harness
already type-checks each turn, so the agent never needs to build itself;
and a mis-issued build---most likely when the model guesses at flags or
targets---scatters stale compiled artifacts (\texttt{.vo}, \texttt{.vos},
\texttt{.glob}) across the tree, corrupting the project's incremental
build state and derailing every later attempt; recovering would then
require a human to clear those files, breaking the otherwise
unattended run.

The two \texttt{Turn} post-hooks run once the turn
completes. \texttt{verify\_proof} is the one described above; besides
type-checking with the Coq kernel, it also runs a coverage check that no
target lemma was dropped or altered. The other, \texttt{iris\_lint},
enforces Iris's coding conventions; we place it as a post-hook rather than
a pre-hook so the LLM can first concentrate on finding a working proof,
with the linter applied afterward to polish it into Iris's conventions.

To see these hooks in action, consider proving a lemma in Iris.
For it, the \texttt{prove\_lemmas} workflow issues
an \texttt{invoke}, the primitive that runs one model interaction. As
Figure~\ref{lst:hhl} shows, \texttt{invoke} takes four arguments: the
\texttt{coq\_project} \emph{workspace} (the run environment---working
directory, model, system prompt, and allowed tools), a \emph{prompt}, and
the lists of \emph{pre-} and \emph{post-hooks}. Here the prompt merely
\emph{names} the target (\texttt{"Prove the lemma " ++ lemma})---not
its statement; the agent reads the source file to recover the statement
and the definitions and prior lemmas in scope, then writes a proof into
the file as a sequence of Coq tactics. Each write is gated by the
pre-hooks, and when the turn ends the \texttt{Turn} post-hooks run; if
every hook passes, \texttt{invoke} returns success and the workflow
records the lemma as proved, otherwise it returns the verifier's error.
The workflow wraps \texttt{invoke} in a feedback-driven retry loop, written
in Figure~\ref{lst:hhl} as a \texttt{while} over three \texttt{var}
bindings: \texttt{proved} (set from \texttt{!r.is\_error}), a \texttt{tries}
counter, and \texttt{feedback}, the verifier's error read from
\texttt{r.result}. On rejection, \texttt{feedback} is appended to the next
\texttt{invoke}'s prompt, so the agent sees what went wrong and revises;
the loop condition \texttt{!proved \&\& tries < 30} keeps retrying until a
proof is accepted or the thirtieth attempt, after which the lemma is
recorded as failed. Aria additionally
resumes the \emph{same session} across retries, carrying the full error
history forward.

\subsection{Syntax and Compilation}
\label{sec:chl:syntax}

An HHL program declares \texttt{hook}s that bind a tool event to a Python
policy function, \texttt{workspace}s that fix the run environment, embedded
Python, and one or more \texttt{workflow}s; execution starts from a
designated entry-point workflow. A workflow orchestrates with ordinary
control flow (\texttt{let}/\texttt{var}, \texttt{if}, \texttt{while},
\texttt{for}, \texttt{match}), may \texttt{call} other workflows, and drives
the model through the \texttt{invoke} primitive, which runs one model
interaction under chosen pre-/post-hooks and returns a \texttt{Response} the
workflow branches on (\texttt{is\_error}, \texttt{result}). Each hook receives the event
payload---a \texttt{File}, \texttt{EditOp}, or \texttt{Command}---and
returns \texttt{Allow}/\texttt{Deny} for pre-hooks or
\texttt{Ok}/\texttt{Err} for post-hooks; its body is ordinary Python, so it
can grep a file, call an external checker, or shell out to the prover.

HHL is compiled, not interpreted: the compiler resolves \texttt{require}
dependencies and lowers the program to Python on a target agent
runtime---currently the Claude Code SDK~\cite{ClaudeCodeSDK}, though,
because it depends only on tool dispatch and pre-/post-tool hook points, it
can be retargeted (e.g.\ to the OpenAI Agents SDK) without changing any hook
or workflow. Table~\ref{tab:hhlmap} gives the mapping: each hook becomes a
Python script wired into the SDK's \texttt{PreToolUse}/\texttt{PostToolUse}
protocol through a generated settings file (a \texttt{Turn} hook, having no
native counterpart, becomes post-turn logic in the driver), the
\texttt{workspace} becomes the run configuration, and each \texttt{invoke}
becomes one \texttt{claude -p} run---a non-interactive (headless) shell
command, so the whole workflow executes as a program with no human at the
keyboard.

\begin{table}[t]
\centering
\caption{How the HHL compiler lowers each construct.}
\label{tab:hhlmap}
\footnotesize
\begin{tabular}{ll}
\toprule
\textbf{HHL construct} & \textbf{Claude Code SDK target} \\
\midrule
\texttt{pre} hook on \texttt{Write}/\texttt{Edit} & \texttt{PreToolUse} (\texttt{Write|Edit}) \\
\texttt{pre} hook on \texttt{Bash}                & \texttt{PreToolUse} (\texttt{Bash}) \\
\texttt{post} hook on \texttt{Write}/\texttt{Edit}& \texttt{PostToolUse} (\texttt{Write|Edit}) \\
\texttt{post} hook on \texttt{Turn}               & driver post-turn logic \\
\texttt{workspace}                                & run configuration \\
\texttt{invoke}                                   & a \texttt{claude -p} run \\
\bottomrule
\end{tabular}
\end{table}

\section{Aria Implementation}
\label{sec:implementation}

Aria runs as a single, long-lived campaign. Once launched on the four
core Iris modules, it proves all 4,257 lemmas autonomously---discovering
the targets, generating and repairing proofs, keeping the project in a
buildable state, and logging every outcome---with no human expert in the
loop: from launch to the last completed proof, nobody inspects a goal,
writes a tactic, or restarts the run by hand. The campaign we report ran
for roughly 380 hours of model time (about 16 days); the only human action it required was
starting it.

\begin{figure}[t]
\centering
\begin{tikzpicture}[
  font=\scriptsize,
  box/.style={rectangle, rounded corners, draw, thick, align=center, inner sep=3pt},
  driver/.style={box, fill=orange!20, minimum width=4.2cm, minimum height=8mm},
  agent/.style={box, fill=orange!8, minimum width=1.9cm, minimum height=10mm},
  harn/.style={box, fill=green!15, minimum height=9mm},
  arr/.style={-{Stealth[length=1.6mm]}, semithick},
  biarr/.style={{Stealth[length=1.6mm]}-{Stealth[length=1.6mm]}, semithick},
  lbl/.style={font=\tiny\itshape, inner sep=1.5pt}
]
\node[driver] (driver) at (0,3.1)
  {\textbf{Driver} (HHL workflow)\\orchestrates the campaign};
\node[agent] (ext) at (-2.9,1.5)
  {\textbf{Extractor}\\unproved\\$\rightarrow$ lemma names};
\node[agent] (prv) at (0,1.5) {\textbf{Prover}\\first attempt};
\node[agent] (fix) at (2.5,1.5) {\textbf{Fixer}\\repairs on error\\(retry $\le 30$)};
\node[harn, minimum width=2.5cm] (exharn) at (-2.9,-0.5)
  {\textbf{Extraction}\\\textbf{harness}\\lemma-count check};
\node[harn, minimum width=5cm] (coqharn) at (1.2,-0.55)
  {\textbf{Coq Harness} (kernel $+$ hooks)\\\texttt{verify\_proof},
   \texttt{iris\_lint}, \texttt{no\_admitted},\\\texttt{safe\_command}, \texttt{timeout}};
\node[agent, minimum width=2.4cm] (pol) at (1.2,-2.4)
  {\textbf{Polish}\\rubric style (retry $\le 2$)};

\draw[biarr] (driver) -- (ext) node[lbl, midway, left]{extract\,/\,names};
\draw[arr]   (driver) -- (prv) node[lbl, midway, right]{lemma};
\draw[biarr] (ext) -- (exharn) node[lbl, midway, right]{validate};
\draw[arr] (prv.south) -- (prv.south |- coqharn.north)
  node[lbl, midway, left]{proof};
\draw[arr] ([xshift=-0.4cm]fix.south |- coqharn.north) -- ([xshift=-0.4cm]fix.south)
  node[lbl, midway, left]{error$+$goal};
\draw[arr] ([xshift=0.4cm]fix.south) -- ([xshift=0.4cm]fix.south |- coqharn.north)
  node[lbl, midway, right]{fixed proof};
\draw[arr] ([xshift=-0.5cm]coqharn.south) -- ([xshift=-0.5cm]coqharn.south |- pol.north)
  node[lbl, midway, left]{accepted};
\draw[arr] ([xshift=0.5cm]pol.north) -- ([xshift=0.5cm]pol.north |- coqharn.south)
  node[lbl, midway, right]{re-check};
\end{tikzpicture}
\caption{Aria's multi-agent proof loop. Each LLM role (Extractor, Prover,
Fixer, Polish) is paired with a harness that validates its output.}
\label{fig:agents}
\end{figure}
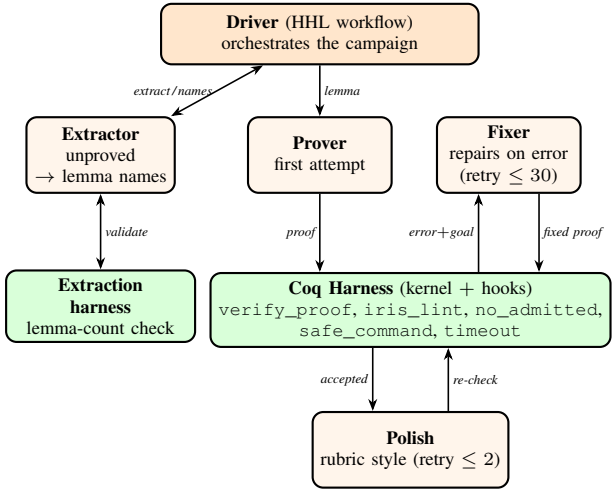

Under the hood, Aria runs as a small team of specialized LLM agents,
coordinated by a \textit{driver} program and kept honest by automatic
checkers (Figure~\ref{fig:agents}). The driver runs the whole campaign.
It first uses an \textit{Extractor} simply to find which lemmas still need
a proof; a quick completeness check (the extraction harness in
Figure~\ref{fig:agents}) confirms that none were missed. The driver then
gives each lemma to a \textit{Prover}, which writes a first proof
attempt and submits it to the Coq harness. If the harness rejects it, the
error message and the still-open goal are handed to a \textit{Fixer},
which revises the proof and resubmits---looping, always with the latest
error and goal in hand, until the harness accepts the proof or an attempt
limit is reached. A \textit{Polish} agent then rewrites the accepted proof
toward the style rubric, with the harness re-checking each rewrite. The
model is never taken at its word: a lemma counts as proved only once the
Coq harness has checked it.

\subsection{Critical Harness Design}

The verification harness is the core of Aria and its only source of trust.
Two checks turn each rejection into usable feedback: a \emph{Coq
verification} check that pins an error to a line and goal, and a
\emph{divergent-tactic} check that catches proofs that never terminate.

\subsubsection{Coq verification}
After each turn the harness takes the single \texttt{.v} file the agent has
just modified and checks it with Coq, one proof step at a time. If the
file compiles cleanly, the turn is accepted. If a step fails, the harness
stops at that point and returns three things: the \textit{line} where the
proof broke, Coq's \textit{error message}, and the \textit{pending goal}
at that line---the proof state Coq was in when the tactic failed,
recovered by replaying the proof up to the failing step.

This is what turns a rejection into usable feedback. A bare ``the proof
failed'' tells the model nothing; an error pinned to a line, together with
the goal that still has to be proved, tells it exactly where it went
wrong and what remains. For instance, on an attempted proof of
$P \wedge Q \rightarrow Q \wedge P$ whose last step wrongly applies the
hypothesis \texttt{HP : P} to the goal \texttt{Q}, the harness reports:

\begin{lstlisting}[basicstyle=\footnotesize\ttfamily, frame=single,
  xleftmargin=1em, breaklines=true, aboveskip=4pt, belowskip=2pt]
(1) failing step (line 4): apply HP.
(2) error message: Unable to unify "P" with "Q".
(3) pending goal:
      HP : P,  HQ : Q
      ----------------
      Q
\end{lstlisting}

\noindent The model now knows the offending tactic and the goal it must
discharge, and can revise accordingly.

\subsubsection{Divergent-tactic check}
Some tactics \textit{diverge}---they neither succeed nor fail, but run
forever, as when
a heavy automation tactic searches without end or a \texttt{repeat} loops
without making progress. Coq itself never comes back---there is no error to
catch, only a hang---so the harness caps each tactic at 300\,s: when a step
exceeds that, the harness stops it and reports a \textit{timeout} at the stuck
step, together with the pending goal just before that tactic. This both tells
the model the step is too expensive to use and keeps one divergent attempt from
stalling the whole campaign.

\subsection{Session (Context) Management}
\label{sec:session-mgmt}

The proof loop is driven by a conversational \textit{session} with the
model, and how that session is reused is central to Aria's effectiveness.
The workflow asks the model to prove each lemma and submits the file to the
harness, retrying up to 30 times on rejection, each time feeding back the
error and the pending goal. The retries are not
independent: \textit{each retry resumes the same session}, so the model
keeps the entire conversation---its earlier attempts and the accumulated
chain of errors and goals---and builds on what it has already tried
instead of repeating dead ends. The
\textit{Polish} agent, by contrast, runs in its own fresh session, sharing
none of the Prover's or Fixer's history: polishing starts from the
accepted proof alone, keeping it independent of---and a fair check on---the
proof-finding effort.

Sessions are reused across lemmas as well, but only briefly---a
deliberate tradeoff. Within a file, a session carries over at most six
consecutive lemmas before it is reset, and every new file begins fresh.
Reuse helps because adjacent lemmas in an Iris file tend to share
definitions and proof ideas, so the model can carry over tactics it has
just found; the window of six is a rule of thumb that assumes such nearby
lemmas are related. Push the reuse further, though, and the session
accumulates a long history that dilutes the relevant context and slows
the model down, making it less likely to land a proof quickly. Resetting
costs little: a fresh session can always re-read whatever it needs from
the source file, so it loses the conversation, not the information.

\section{Experiment}
\label{sec:experiment}

Our implementation of the agents is available in an anonymized repository:
\url{https://anonymous.4open.science/r/CoqProver-Code-57BB}. The experiments
were conducted on a Claude Code 20$\times$ subscription rather than the
metered API, which significantly reduced cost.

\subsection{Experimental Setup}

\textbf{Benchmarks.} We evaluate Aria on four independent bodies of
proofs---three in Coq and one in Lean. The primary target is the
\textit{Iris core}---the four core
modules of the Iris Coq
development~\cite{IrisRepo}, comprising \textit{every} lemma whose original,
expert-written proof terminates in \texttt{Qed}: 4,257 lemmas across 113
files. We deliberately include the entire population rather than a sample,
so our results characterize the complete modules. To test whether the
approach carries beyond the Iris core, we add three further benchmarks: the
\textit{Rust library} Coq verification in the RustBelt
project~\cite{LambdaRustRepo} (217
lemmas); \textit{reglang}, a Coq library of regular-language
theory~\cite{ReglangRepo} that is the
CoqStoq project on which prior LLM provers fare
worst~\cite{Thompson2025Rango}; and \textit{iris-lean}~\cite{IrisLean},
three not-yet-ported files of the in-progress Lean~4 port of Iris (72
lemmas), whose proofs are checked by the Lean kernel rather than Coq's.

\textbf{Proof-regeneration protocol.} For each target lemma we remove its
original proof body, leaving the lemma statement intact, and task the
agent with reconstructing a proof from scratch. The remainder of the
source file---surrounding definitions and notations---together with all
imported dependencies remains
available as context; only the proof under test is removed.

\textbf{Resource restrictions.} So the agent reconstructs each proof
rather than retrieving it, we run it in a sandbox with no network and no
version control: it can neither browse the web nor recover the removed
proofs from \texttt{git} history. Its only resources are the on-disk
\texttt{.v} files---with the target proofs removed---and the dependent
libraries installed via \texttt{opam}. Any proof it produces is therefore
built from the lemma statement and its in-scope context, not copied from
the reference proof.

\textbf{Models.} Unless stated otherwise, all results use a single
backend---Claude Opus~4.7---with no other model involved. Aria's model
layer is nonetheless backend-agnostic.
We examine how the choice of model affects the outcome separately in
Section~\ref{sec:exp:models}.

\textbf{Metrics.} For each lemma we record whether it was \textit{proved},
the number of \textit{retries} it took, and the \textit{time} to that proof.
The reported time is \emph{model time}---the wall-clock time the LLM spends
reasoning and generating the proof.
A retry is one in-session repair the agent makes after the
harness rejects a candidate, so a lemma's retry count is its number of
failed attempts and a \textit{first-attempt} success is a retry count of
zero. The agent retries up to a cap of \textbf{30} per lemma; if the cap
is reached without an accepted proof, the lemma is recorded as a failure.

\subsection{Iris Core}
\label{sec:exp:iris}

The four core modules of Iris, from most abstract to most concrete:
(1)~\texttt{algebra} defines the resource structures that track ownership;
(2)~\texttt{bi} defines the logic's connectives, such as the separating
conjunction; (3)~\texttt{base\_logic} adds invariants and other mechanisms
for shared state; and (4)~\texttt{program\_logic} adds the rules relating
code to its specification.

Table~\ref{tab:results} summarizes the outcome of running Aria over
the Iris Coq development. Aria proved \textbf{4,257 distinct lemmas}
across 113 source files with zero Coq expert intervention and no failures. The vast majority of lemmas---79.2\% (3,373 of 4,257)---were solved on the
agent's first attempt, with no retry needed, and the mean retry count per
lemma was only 0.51, indicating that the
LLM~+~harness loop converges quickly for the overwhelming majority of
proof obligations. This high first-attempt rate is not a sign of
memorization: apart from trivial one-line lemmas, whose proof is
essentially forced, the proofs Aria generates differ from the upstream Iris
proofs---it constructs them rather than recalling them.
The most any lemma required was 28 retries---below the
cap of 30---so no lemma ever exhausted the retry budget, consistent with
the absence of failures. The full run consumed roughly 380 hours of model
compute, averaging 321 seconds per lemma; per-lemma cost ranged from a few
seconds for simple structural lemmas to over 4.7 hours for the single most
demanding obligation.

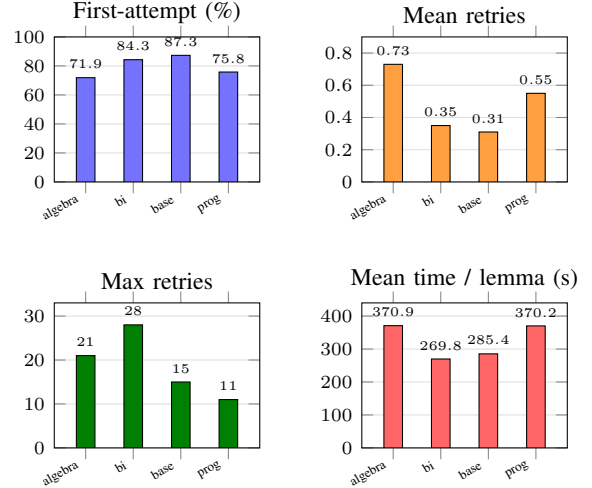
\begin{figure}[t]
\centering
\begin{tikzpicture}
\begin{groupplot}[
  group style={group size=2 by 2, horizontal sep=1.35cm,
    vertical sep=1.6cm},
  width=4.3cm, height=3.5cm,
  ybar, /pgf/bar width=7pt,
  symbolic x coords={algebra,bi,base,prog},
  xtick=data,
  x tick label style={font=\tiny, rotate=30, anchor=east},
  y tick label style={font=\scriptsize},
  title style={font=\small, yshift=-1ex},
  enlarge x limits=0.22,
  ymajorgrids, grid style={gray!25},
  nodes near coords, nodes near coords style={font=\tiny},
  every node near coord/.append style={anchor=south},
]
\nextgroupplot[title={First-attempt (\%)}, ymin=0, ymax=100]
\addplot[fill=blue!55] coordinates
  {(algebra,71.9) (bi,84.3) (base,87.3) (prog,75.8)};
\nextgroupplot[title={Mean retries}, ymin=0, ymax=0.9]
\addplot[fill=orange!75] coordinates
  {(algebra,0.73) (bi,0.35) (base,0.31) (prog,0.55)};
\nextgroupplot[title={Max retries}, ymin=0, ymax=33]
\addplot[fill=green!50!black] coordinates
  {(algebra,21) (bi,28) (base,15) (prog,11)};
\nextgroupplot[title={Mean time / lemma (s)}, ymin=0, ymax=440]
\addplot[fill=red!60] coordinates
  {(algebra,370.9) (bi,269.8) (base,285.4) (prog,370.2)};
\end{groupplot}
\end{tikzpicture}
\caption{Per-module results on the Iris core (\texttt{base} =
\texttt{base\_logic}, \texttt{prog} = \texttt{program\_logic}). Over the
$4{,}123$ lemmas ($96.9\%$) with module-qualified log paths; the
remaining $134$ have ambiguous file names (e.g.\ \texttt{big\_op.v}
occurs in two modules) and are omitted.}
\label{fig:iris-modules}
\end{figure}

Figure~\ref{fig:iris-modules} breaks these results down by module. The
profile is uniform across all four: every module is solved completely,
with first-attempt rates between $72$ and $87\%$ and well under one retry
per lemma on average. The \texttt{algebra} module---the resource-algebra
layer---is the most demanding, with the lowest first-attempt rate
($71.9\%$) and the highest mean retries and time per lemma, while the more
derivative \texttt{bi} and \texttt{base\_logic} layers converge fastest.
Notably, retries and time do not track each other: in
Figure~\ref{fig:iris-modules}, \texttt{program\_logic} has the
\emph{fewest} maximum retries (11) yet nearly the highest mean time per
lemma, whereas \texttt{bi} has the \emph{most} (28) yet the lowest mean
time. This indicates that the reasoning time of each attempt varies with
the logical complexity of the lemma, independent of the number of
retries.

\begin{table}[t]
\centering
\caption{Aria results on the core modules of the Iris Coq
development (distinct lemmas, duplicate re-runs removed).}
\label{tab:results}
\begin{tabular}{lr}
\toprule
\textbf{Metric} & \textbf{Value} \\
\midrule
Distinct lemmas proved        & 4,257 \\
Source files covered          & 113 \\
Failures                      & 0 \\
First-try success (0 retries) & 3,373 (79.2\%) \\
Mean retries per lemma        & 0.51 \\
\midrule
Total model time              & 379.7 h (15.8 d) \\
Mean time per lemma           & 321.1 s \\
Min / max time per lemma      & 8.7 s / 17{,}103 s \\
\bottomrule
\end{tabular}
\end{table}

\subsection{Rust Standard Libraries}
\label{sec:exp:rustbelt}

To test whether Aria's competence extends beyond the Iris core to the
verified software built on it, we ran the same agent and hardened harness
on \textbf{RustBelt}, the Iris-based verification of Rust that gives the
first formal soundness proof for Rust's ownership-based type system,
including the \texttt{unsafe} code in its standard
library~\cite{Jung2018RustBelt}. We target Rust's
libraries for concurrency and interior mutability
(\texttt{Arc}, \texttt{Rc}, \texttt{Mutex}, \texttt{RwLock},
\texttt{RefCell}, \texttt{Cell}, \ldots), which build
with a \texttt{make}-based toolchain rather than Iris's
dune setup, but are otherwise driven by the identical protocol: strip each
lemma's proof, let Aria derive a new one, and accept it only when the Coq
kernel checks it.

Aria proved \textbf{217 lemmas with zero failures} across 25 files,
\textbf{73\%} on the first
attempt and \textbf{86\%} within a single retry; the hardest cases---such
as the drop and clone proofs for \texttt{Arc} and the reader/writer guards
of \texttt{RwLock}---took up to 13 retries. That Aria carries
over---the same zero-failure outcome and first-try-dominated retry
profile---from the Iris core to a downstream, real-world Rust-safety
development is evidence the result is not an artifact of overfitting to a
single corpus.

\begin{table}[t]
\centering
\caption{Aria on the Rust libraries verified in RustBelt.}
\label{tab:rustbelt}
\footnotesize
\begin{tabular}{lr@{\hskip 2em}lr}
\toprule
\textbf{Library} & \textbf{\#Lemmas} & \textbf{Library} & \textbf{\#Lemmas} \\
\midrule
\texttt{RwLock}  & 36 & \texttt{Rc}     & 30 \\
\texttt{RefCell} & 35 & \texttt{Mutex}  & 20 \\
\texttt{Arc}     & 33 & \texttt{Vec}    & 17 \\
\texttt{Cell}    & 35 & \texttt{Option} & 11 \\
\bottomrule
\end{tabular}
\end{table}

\subsection{reglang}
\label{sec:exp:reglang}

reglang is a Coq library of regular-language theory---finite automata,
regular expressions, and their equivalences. Unlike the Iris core and RustBelt, its
proofs involve neither separation logic nor concurrency but instead
underpin string-type verification in programming languages; we include it
because, among the twelve projects in the CoqStoq
benchmark~\cite{Thompson2025Rango}---which draws on the
CoqGym~\cite{Yang2019CoqGym} projects that compile under Coq~8.18,
together with the CompCert verified C compiler and several actively
maintained Coq-Community libraries---reglang is the
one on which prior LLM provers such as Rango fare \emph{worst}, proving
barely one in eight, which makes it a direct comparison against prior work
on its hardest case. We apply the identical protocol: strip each lemma's
proof and accept a new one only when the Coq kernel checks it.

Aria proves \textbf{all 318 lemmas with zero failures}, \textbf{75.5\%}
on the first attempt and a mean of \textbf{0.38 retries} per lemma---the
lowest retry rate of the three Coq benchmarks, consistent with reglang being
the least demanding target. No proof required more than six retries. The
averages hide one striking outlier, however: \texttt{nfa\_ofP}---the
correctness of a two-way finite-automaton construction---was the
\emph{single longest-running proof in the entire study}, taking roughly
\textbf{5.8 hours} and four retries to land (longer even than
the hardest Iris lemma, at 4.7 hours). Where prior LLM provers clear
barely one in eight of these proofs, Aria proves them all.

\subsection{Cross-Benchmark Comparison}
\label{sec:exp:comparison}

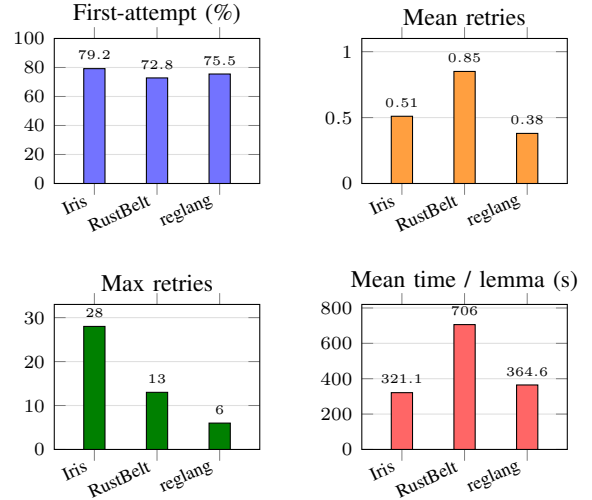
\begin{figure}[t]
\centering
\begin{tikzpicture}
\begin{groupplot}[
  group style={group size=2 by 2, horizontal sep=1.35cm,
    vertical sep=1.6cm},
  width=4.3cm, height=3.5cm,
  ybar, /pgf/bar width=8pt,
  symbolic x coords={Iris,RustBelt,reglang},
  xtick=data,
  x tick label style={font=\scriptsize, rotate=25, anchor=east},
  y tick label style={font=\scriptsize},
  title style={font=\small, yshift=-1ex},
  enlarge x limits=0.32,
  ymajorgrids, grid style={gray!25},
  nodes near coords, nodes near coords style={font=\tiny},
  every node near coord/.append style={anchor=south},
]
\nextgroupplot[title={First-attempt (\%)}, ymin=0, ymax=100]
\addplot[fill=blue!55] coordinates
  {(Iris,79.2) (RustBelt,72.8) (reglang,75.5)};
\nextgroupplot[title={Mean retries}, ymin=0, ymax=1.1]
\addplot[fill=orange!75] coordinates
  {(Iris,0.51) (RustBelt,0.85) (reglang,0.38)};
\nextgroupplot[title={Max retries}, ymin=0, ymax=33]
\addplot[fill=green!50!black] coordinates
  {(Iris,28) (RustBelt,13) (reglang,6)};
\nextgroupplot[title={Mean time / lemma (s)}, ymin=0, ymax=820]
\addplot[fill=red!60] coordinates
  {(Iris,321.1) (RustBelt,706) (reglang,364.6)};
\end{groupplot}
\end{tikzpicture}
\caption{Cross-benchmark comparison of Aria on the Iris core (4{,}257
lemmas), RustBelt (217), and reglang (318).}
\label{fig:comparison}
\end{figure}

Figure~\ref{fig:comparison} compares Aria across the three Coq developments on
four per-lemma metrics: the first-attempt success rate, the mean and
maximum number of retries, and the mean model time per lemma. The
profile is consistent across all three: a first-attempt rate in the
$73$--$79\%$ range and a mean of well under one retry per lemma, with the
Iris core---the largest and hardest target---posting both the highest
first-attempt rate and the longest retry tail (up to 28). The
RustBelt proofs are the most expensive on average
($706$\,s per lemma, roughly double the others), reflecting their greater
length rather than a higher failure rate.

\subsection{iris-lean: Lemmas with No Prior Proof}
\label{sec:exp:lean}

Our final benchmark pushes on two axes at once: a different proof
assistant (Lean), and lemmas whose Lean proofs have never existed and so
could not have been learned by the state-of-the-art model. Because Iris is so widely used, an effort is
underway to port it to Lean~4---iris-lean~\cite{IrisLean}---but the port is
far from finished, and a public tracker~\cite{IrisLeanTracker} shows which
files have not yet been added. We pick three of these not-yet-ported files (\texttt{Function},
\texttt{Mra}, and \texttt{UFrac}), and task Aria with supplying the
proofs. Soundness is now decided by the Lean kernel rather than Coq's---the
same harness design retargeted to a different verifier.

Aria proves all \textbf{72 lemmas with zero failures}, \textbf{90.3\%} on
the first attempt and a mean of \textbf{0.10 retries}---no proof needed
more than one retry---in under \textbf{two hours}, the cleanest profile of
any benchmark in this study. It also shows the harness~+~code~agent
approach is not tied to Coq: the same design works with the Lean kernel as
the verifier.

A direct comparison is instructive. On the three corresponding modules in
upstream Coq Iris (\texttt{functions.v}, \texttt{mra.v}, \texttt{ufrac.v};
49 lemmas), Aria proves every lemma but at a $75.5\%$ first-attempt rate,
up to four retries, and $149$\,s per lemma. The Lean port states the same
algebra as more, smaller lemmas ($72$ vs.\ $49$), and on these finer
obligations Aria is markedly cleaner: $90.3\%$ first-attempt, never more
than one retry, and $97$\,s per lemma. The lesson is
methodological---splitting a development into smaller lemmas gives the agent
simpler obligations, raising the first-attempt rate and lowering per-lemma
time.

\subsection{Private vs.\ Open-Source Models}
\label{sec:exp:models}

\begin{figure}[t]
\centering
\begin{tikzpicture}
\begin{axis}[
  width=\columnwidth, height=4.8cm,
  ybar, /pgf/bar width=9pt,
  symbolic x coords={A,B,C,D},
  xtick={A,B,C,D},
  xticklabels={1st-attempt (\%), mean retries, max retries,
    time/lemma (s)},
  x tick label style={font=\scriptsize, rotate=12, anchor=east},
  axis y line=none, ymin=0, ymax=1.28,
  enlarge x limits=0.18,
  legend style={font=\scriptsize, at={(0.5,1.03)}, anchor=south,
    legend columns=2, draw=none, fill=none},
  point meta=explicit symbolic,
  nodes near coords, nodes near coords style={font=\tiny, anchor=south},
]
\addplot[fill=blue!55] coordinates
  {(A,1.0) [72.5] (B,0.226) [0.40] (C,0.125) [3] (D,0.135) [169]};
\addplot[fill=orange!75] coordinates
  {(A,0.759) [55.0] (B,1.0) [1.77] (C,1.0) [24] (D,1.0) [1253]};
\legend{Opus~4.7, Kimi~K2.6}
\end{axis}
\end{tikzpicture}
\caption{Private (Claude Opus~4.7) vs.\ open-source (Kimi~K2.6).}
\label{fig:models}
\end{figure}
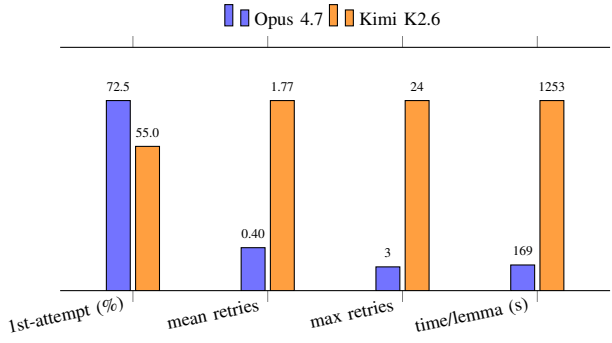

To compare the effectiveness and efficiency of a private and an
open-source model, we re-ran a fixed benchmark---the 40
lemmas of
\texttt{iris/algebra/lib/frac\_auth.v}---under two models, holding the
same Claude Code agent, harness, workflows, and prompts: the private
Claude Opus~4.7 and
the open-source Kimi~K2.6, served through Ollama Cloud behind an
Anthropic-compatible API. Figure~\ref{fig:models} reports the outcome.
On this subset both models proved \textit{all} 40 lemmas, so an
open-source model is clearly usable for the task; the visible difference
is efficiency. The private model solved 72.5\% of the lemmas on the first
attempt, never needed more than three retries, and finished all 40 lemmas
in under two hours; the open model solved 55\% on the first attempt, occasionally
needed many retries (up to 24), and took roughly seven times as long. We
additionally ran DeepSeek~V4~Pro, also through Ollama Cloud, under the
identical setup (not shown in Figure~\ref{fig:models}); its results closely
tracked Kimi's---all 40 lemmas proved with a comparable retry profile---but
at roughly two to three times the token consumption.
The picture changes on harder proofs, however. These 40 lemmas have short
proofs, which is precisely why the open model keeps up; across the wider
library we observed a genuine \textit{capability} gap on long proofs.
Lemmas whose proofs run beyond roughly fifty lines frequently defeated the
open model even after it exhausted the 30-retry budget; Opus~4.7, by
contrast, proved them, reliably constructing and maintaining correct
proofs of more than two hundred lines.

\subsection{Rubric Criteria for Proof Polishing}
\label{sec:exp:polishing}

To illustrate the stylistic differences between human-written and
LLM-generated proofs, we compare two correct proofs of the same proof
obligation: \texttt{discrete\_cmra\_mixin} (a \texttt{Definition} that, like
a lemma, is built by a proof and checked by the Coq kernel).

\medskip
\noindent\textbf{Human proof.}
The expert-written proof is concise, relying on advanced tactic
combinators and domain intuition to dispatch goals efficiently.

\begin{lstlisting}[style=coqstyle]
Definition discrete_cmra_mixin : CmraMixin A.
Proof.
  destruct ra_mix; split; try done.
  - intros x; split; first done.
    by move=> /(_ 0@$_i$@).
  - intros n x y1 y2 ??;
    by exists y1, y2.
Qed.
\end{lstlisting}

\noindent\textbf{LLM proof (first stage).}
As first generated, the LLM proof is more verbose, addressing each
subgoal explicitly rather than relying on compact tactic idioms.

\begin{lstlisting}[style=coqstyle]
Proof.
  destruct ra_mix. split.
  - intros x n. apply ra_op_proper0.
  - intros n x y cx.
    apply ra_core_proper0.
  - intros n. apply ra_validN_proper0.
  - intros x. split;
    [done|intros Hv; apply (Hv 0@$_i$@)].
  - intros n n' x ? ?. done.
  - done.
  - done.
  - done.
  - done.
  - done.
  - done.
  - intros n x y1 y2 _ Hx.
    exists y1, y2. split_and!; done.
Qed.
\end{lstlisting}

Both proofs are accepted by Coq, yet they differ in style. The human
proof uses \texttt{try done} to clear the routine subgoals at once,
leaving a compact body; the first-stage LLM proof instead enumerates
every subgoal, including a long run of bare \texttt{done}---correct, but
verbose and brittle.

Correctness, however, is the \textit{only} hard constraint the Coq
harness enforces; proof \textit{style} is not. To improve style we add a
second, \textit{soft} harness: a set of \textit{rubric criteria} that a
good proof should satisfy:
\begin{itemize}
  \item \textit{Conciseness}: collapse routine, repeated closers and drop
    no-op steps.
  \item \textit{Transparency}: keep each subgoal a legible step---no
    monolithic combinator blobs (e.g.\ a long \texttt{first [\ldots]}).
  \item \textit{Robustness}: avoid dependence on the number, order, or
    auto-generated names of subgoals.
  \item \textit{Idiomatic style}: follow the library's conventions and
    tactic idioms.
  \item \textit{Reuse}: invoke existing lemmas rather than re-deriving
    them.
\end{itemize}
Aria applies these in a second stage. A prover agent first produces a
proof the Coq harness accepts; a separate \textit{polish} agent---started
fresh, sharing none of the prover's context---is then handed that proof.
The soft harness scores it against each criterion and returns feedback
only for the ones it fails; the polish agent rewrites accordingly, and the
Coq harness re-checks every rewrite so polishing can never break
correctness. This repeats until all criteria pass or a cap of two
polishing rounds is reached---if some criteria still fail after that, the
already-correct proof is kept as is. Running this stage on the proof above
yields:

\begin{lstlisting}[style=coqstyle]
Definition discrete_cmra_mixin : CmraMixin A.
Proof.
  destruct ra_mix; split; try done.
  - intros x. split; [done|by intros Hv; apply (Hv 0@$_i$@)].
  - intros n x y1 y2 _ Hx. by exists y1, y2.
Qed.
\end{lstlisting}

The polished proof recovers the expert style: \texttt{try done} collapses
the routine cases, and only the two substantive obligations remain as
explicit bullets---matching the human proof in length and structure while
staying fully machine-checked.

\section{Discussion}
\label{sec:discussion}

The deeper lesson of this work is not about Iris in particular but about
\emph{where the proof strategy should live}. Reading the agents' logs, the
striking thing is how much of each proof is strategic decision-making that
the model carries out entirely on its own.

\textit{The agent decides what to do next.} A code agent follows no fixed
proof procedure; at every step it uses the LLM, and nothing else, to choose
the proof's direction---which prior lemma or hypothesis to apply, how to
break a goal into subgoals and in what order, when to search the source for
a definition it needs, which tactic to try, and when to abandon a line and
backtrack. A recurring pattern
across the logs is that the agent first looks for \emph{similar} lemmas
already in scope and mimics their proofs, adapting a nearby, structurally
related argument rather than building each one from first principles. This
is a strikingly human-like way to prove---the same instinct an expert
relies on when reaching for an analogous lemma instead of starting over.

\textit{These are exactly the decisions prior systems hard-wire.} Where a
trained premise selector, a proof-tree search, and a divide-and-conquer
module each fix one choice in advance, our method delegates all three to
the agent's judgment---no premise model, no search policy, no decomposition
heuristic. Because these decisions are made \emph{in context} and \emph{per
lemma} rather than imposed uniformly by an external policy, a deliberately
coarse loop of agent plus verifier clears proofs that defeated systems
built around elaborate search.

\section{Related Work}
\label{sec:related}

We review prior automation along a single axis---how much of the proving
it hands to a large language model---running from classic symbolic methods,
through learned tactic search, to the model-in-the-loop agents closest to
ours.

\textit{Classic symbolic automation.} \textit{Hammers} discharge a goal by
translating it to a first-order or SMT problem and reconstructing the
result---CoqHammer~\cite{Czajka2018CoqHammer} and
Sledgehammer~\cite{Bohme2010Sledgehammer}, with Thor~\cite{Jiang2022Thor}
adding a language model for premise selection. They are effective on
first-order goals---CoqHammer closes roughly 40\% of the Coq standard
library push-button---but a poor fit for the higher-order, step-indexed,
separation-logic obligations that dominate Iris. Within Iris, the Iris
Proof Mode and MoSeL~\cite{Krebbers2017IPM, Krebbers2018MoSeL} and
Diaframe~\cite{Mulder2022Diaframe} cut proof effort but are
expert-engineered, and non-routine proofs still demand human guidance.
Apart from Thor's learned premise selection, these methods involve no
machine learning.

\textit{Tactic prediction and proof search.} A decade of work learned to
predict tactics from corpora of human proofs and combine them with search
over the proof tree---ASTactic on CoqGym~\cite{Yang2019CoqGym},
Proverbot9001~\cite{SanchezStern2020Proverbot},
Tactician~\cite{Blaauwbroek2020Tactician},
Graph2Tac~\cite{Blaauwbroek2024Graph2Tac}, the goal-adaptive
Gpass~\cite{Chen2025Gpass}, and the diversity-driven ensemble
Diva~\cite{First2022Diva}---the difficulty lying in premise selection and
efficient search. A parallel line invests in search itself: HyperTree Proof
Search~\cite{Lample2022HyperTree} couples a neural prover with
AlphaZero-style search, setting records on Metamath and Lean. These systems
treat proving as guided search; their effort goes into the search policy
rather than the verification contract around the model.

\textit{LLM-based proof generation.} LLMs shifted the emphasis toward
generating whole proofs or large fragments, often with premise retrieval
and feedback-driven repair: Baldur~\cite{First2023Baldur} and
Draft-Sketch-Prove~\cite{Jiang2023DSP} for Isabelle,
LeanDojo~\cite{Yang2023LeanDojo} and
DeepSeek-Prover~\cite{Xin2024DeepSeekProver} for Lean,
GPT-f~\cite{Polu2020GPTf} for Metamath, PALM~\cite{Lu2024PALM},
Rango~\cite{Thompson2025Rango}, COPRA~\cite{Thakur2024COPRA}, and the
divide-and-conquer Cobblestone~\cite{Kasibatla2026Cobblestone} for Coq, and
proof-oriented programming in F$^\star$~\cite{Chakraborty2025FStar}. As
Table~\ref{tab:priorcoq} summarizes, these are fully automatic but
\textit{partial}---the strongest prove roughly 12--48\% of their
benchmarks---and evaluate on general-purpose, sequential developments such
as CompCert. Closest to our setting, Cobblestone also checks every
candidate with the Coq kernel, yet reaches 48\% fully automatically (58\%
with oracle guidance) on general Coq. Aria differs on both axes: it targets
the higher-order, concurrent separation logic of Iris and proves its four
core modules completely.

\textit{Verifier-in-the-loop agents and harnesses.} Pairing a model with a
checker and feeding back its errors is by now common---for proof
repair~\cite{First2023Baldur} and in-context proof
agents~\cite{Thakur2024COPRA}, and more broadly in autonomous agents for
software tasks such as program repair~\cite{Bouzenia2025RepairAgent}. Our
contribution is twofold. First, we make the harness a first-class,
declarative artifact: HHL lets the policy governing the agent be written,
audited, and retargeted across agent runtimes. Second, we show that a sound
verifier---the prover as an infallible oracle~\cite{First2022Diva}---is
necessary but not sufficient: when an agent can rewrite the proof file
freely, it can satisfy the verifier while silently dropping or weakening the
target lemma, leaving the file type-checking yet the task undone. The
harness must therefore also enforce \textit{completeness}---checking that
the target lemma is still present, unweakened, and actually proved. This is
easy to overlook once an agent edits files freely.

\section{Conclusion}
\label{sec:conclusion}

We presented Aria, which proves theorems fully automatically by pairing a
general LLM code agent with a verification harness. Running unattended, it
proved all 4,257 lemmas of the four core Iris modules, 217 RustBelt
lemmas, and all 318 reglang lemmas---with zero Coq expert intervention and
no failures, the first time an entire expert-grade concurrent
separation-logic library has been proved automatically and in full. The
recipe is deliberately coarse: the agent makes every proof-strategy
decision while the harness, captured as reusable infrastructure in HHL,
keeps it honest. A natural next step is to apply the same
agent-plus-harness recipe upstream, to generating the formal
specifications that proving presupposes.

\bibliographystyle{IEEEtran}
\bibliography{references}

\end{document}